# On the Quest for Standard Model Cold Dark Matter

Bikash Sinha

Variable Energy Cyclotron Centre, 1/AF Bidhannagar,

Kolkata 700064, India

*bikash@vecc.gov.in*

The possibility that the relics of quark hadron phase transition in the microsecond old universe, the quark nuggets, may well be reasonable candidates for cold dark matter is critically examined.

## 1.    Introduction:

Over the years abundant evidence has been accumulated indicating the presence of large quantities of unseen matter surrounding normal galaxies including our own [1,2,3]. The nature of this 'dark matter' remains unknown except it cannot be normal, such as stars, gas or dust.

Axions, massive neutrinos, other kind of yet unknown exotic particles, even other weakly interacting massive particles, referred to usually as WIMPs have been proposed as candidates of dark matter but not yet experimentally observed so far. Alcock et al. [4] and Aubourg et al. [5] (using gravitational microlensing method suggested by Paczynski [6]) discovered the existence of this dark matter. Alcock et al. [4] suggested that the dark matter can be explained by normal matter, known collectively as Massive Compact Holo Objects (MACHOs) [7]. They went on to speculate that MACHOs might be brown dwarfs or Jupiter like objects, neutron stars, even black holes. It has been argued by us [8,9,10,11] quite exhaustively, that MACHOs cannot be any of the above [8] but a natural explanation will be that the MACHOs are the relics from the putative cosmic phase transition from quark to hadrons about a microsecond after the Big Bang. MACHOs, it is argued, can be the quark nuggets which survived from that primordial epoch. These relic quark nuggets, it is entirely plausible are made of strange matter, the true ground state of QCD [12].

This assertion has acquired further credibility from the recent experimental observations of the Bose Institute group [13,14] engaged in studying cosmic strangelets at mountain altitude. Although heavy ion collision is unlikely to produce strangelets, for the cosmic phase transition scenario however, strange quark nuggets in the form of strangelets hurtling down through the cosmos will tend to pick up mass [13] from the atmospheric atoms as it reaches the earth. The analysis of Banerjee et al. [13] indicates that a strangelet with an initial mass ~ 64 amu and charge ~ 2 (typically for strangelet Z/A ≪ 1) acquires mass as it passes through the atmosphere, evolves to a mass ~ 340 amu or so, at the end of its cosmic journey. This is at an altitude of ~ 3.6 km above the sea level, typically the Himalayan mountain region in India such as Darjeeling. The collected data, just mentioned, suggest the interpretation of exotic cosmic ray events of very small Z/A, arising from the Strange Quark Matter droplets.

## 2. Cosmic Quark Hadron Phase transition

Armed with the recent observations [13,14] of strangelets at mountain top and the very first observations of MACHO collaboration at Mount Stromlo by Alcock et al. [4] and by Aubourg et al. [5] of EROS collaboration at La Silla, Chile, it is proposed to go through a critical analysis of the origin of Strange Quark Matter (SQM) and their survival through the cosmological time scale.

There are two central issues: one, the very formation of quark nuggets and second, their survival in a cosmic scale. Then, there is very vital but relatively straightforward issue of quark nuggets being made of strange quark matter (SQM).

Around a few microsecond after the big bang the universe went through a phase transition from a universe of quarks, photons and leptons to a universe of hadrons, photons and leptons. The temperature was around $T_c \approx$ (150-200) MeV. As per the wisdom of lattice gauge calculation the universe at that primordial epoch, went through a rapid cross over from quarks to hadrons [15,17].

## 3. Cosmic Phase Transition from Quark to Hadrons

The universe, about a microsecond after big bang, consisted of quarks, gluons, leptons and photons. After reaching a temperature of QCD energy scale T~ 150 MeV, the plasma of quarks and gluons, usually referred to as QGP went through a phase transition to the hadrons.

It is conventionally assumed that the baryon asymmetry $\eta_B = (n_B - n_{\bar{B}})/\gamma$ at that epoch of phase transition is the same as it is in today's universe ~ $10^{-10}$. There are reasonably straightforward arguments [12,15,16] however to show that baryon number density at that epoch dominated by quarks and gluons can be much larger. The consequence of such a possibility is discussed here.

For $\eta_B$ ~ $10^{-10}$, the wisdom of lattice will lead the universe to cross over to hadrons, (Fig 1) thereby erasing all the memories of the universe at earlier times with no possible relic of the QCD phase transition at all times.

Witten [12] and others [8,17] have argued that a first order phase transition is plausible with a "small" supercooling which would imply that the transition occurs effectively at a temperature at which most of the latent heat between the two phases still remains, so that phase coexistence can be established after nucleation. In a recent private communication Witten [18] further asserted that if $\eta_B \approx n_B/\gamma$ remains $10^{-10}$ as it is in the current universe, then supercooling is implausible. However, he also points out [18] that if the baryon to photon ratio is not small during the QCD phase transition and become small because of some phenomena at later times, then supercooling is plausible in principle.

This is the central issue, the relevance of baryon asymmetry at that primordial epoch.

One of the more compelling scenarios of baryogenesis is based on its generation from leptogenesis through topological sphaleron transitions occurring around the electroweak transition temperature Leptogenes, in its turn, occurs through out-of-equilibrium decays of heavy right-handed neutrinos which occur naturally within a seesaw mechanism, leading to Majorana masses for neutrinos (as well as neutrino oscillation parameters) within observable ranges. Fermions with only Majorana masses and no Dirac masses ( Majorana fermions) are charge self-conjugate spin ½ particles for any global U(1) charge. If this U(1) charge is associated with lepton number, then the charge self-conjugate property automatically implies that Majorana mass terms violate lepton number. Thus, it is this supposedly Majorana nature of neutrinos (even if they have a Dirac component as well) which lies at the heart of the incipient lepton number violation. The positive aspect of this mechanism of leptogenesis-induced baryogenesis is that one obtains a numerical result close to the observed baryon-photon ratio of $\mathcal{O}(10^{-10} – 10^{-9})$ without any fine tuning.

The resolution of the issue as to whether neutrinos are predominantly Majorana fermions, as happens to be the common prejudice currently, is to be decided by the currently ongoing experiments on neutrinoless double beta decay. If, contrary to extant belief, such experiments happen to yield null results, and neutrinos are confirmed to be Dirac fermions, this scenario of baryogenesis loses its prime attraction, entailing unsavoury fine tuning.

Given such a volatile situation, alternative scenarios of baryogenesis cannot be ruled out. Prominent among these is the nonthermal Affleck-Dine mechanism [19], This is based on out-of-equilibrium decays of heavy s quarks and s leptons (which, respectively, carry baryon and lepton number) within a supersymmetric framework. When supersymmetry is unbroken, the scalar potential for s quarks and s leptons has flat directions in scalar field space (directions along which the potential does not change). The existence of such flat directions permits s quark and s lepton fields to slide freely and hence acquire very large vacuum values. When supersymmetry breaks down, these flat directions no longer remain flat, but extremize the potential at the large expectation values acquired by the s quark and s lepton fields. These fields are now endowed with masses of order these large vacuum values.

As the universe expands, the Hubble parameter assumes a magnitude close to these masses, the heavy B and L-endowed scalars oscillate and decay into ordinary quarks and leptons, violating B and L. A large baryon asymmetry can be produced at temperatures ($M_Z$) (or slightly higher), without any fine tuning.

The Affleck-Dine mechanism [19] has the potential to produce a baryon asymmetry of (1) without requiring superhigh temperatures. However, the observed baryon asymmetry of ($10^{-10}$) at CMB temperatures needs to emerge naturally from such a scenario. This is what is achieved through a 'little inflation' of about 7 e-folding occurring at a lower temperature which may be identified with the QCD phase transition thought of as a first order phase transition [15]. Such an inflation naturally dilutes the baryon photon ratio to the observed range, even though the baryon potential before the first order phase transition may have been high (of $\mathcal{O}(1)$ in photon units).Comparing this "little inflation" with the more standard Guth's inflationary model [20] one finds that the pattern of entropy variation in the two cases are very different. In the standard inflationary model [20] the entropy is conserved during exponential expansion, and increases, due to reheating when bubbles collide, at the end of the transition. In the "little inflation" scenario on the other hand, the entropy is constantly increasing during the quark hadron phase transition. In more general terms the scale factor R gets multiplied by $\sim 10^4$ where T only decreases by only ~70 MeV and entropy increases by $3 \times 10^9$ times [16].

The increase in entropy changes $n_q/n_\gamma$ radically. Whereas, the quark number density decreases as $n_q \sim R^{-3}$ and $n_\gamma \sim T^3$, the ratio $n_q/n_\gamma \sim (RT)^{-3}$, is proportional to the inverse of entropy. This immediately implies that if at the end of the transition $n_q/n_\gamma \sim 10^{-10}$ as dictated by primordial nucleosynthesis at $T \approx T_c$, $n_B/n_\gamma \sim (1)$. This result is in sharp contrast with normal adiabatic expansion in which the baryon asymmetry will not change [16].

The possibility and the criterion of a mini- inflationary epoch can be demonstrated in a simple way within the Friedman model of a spatially flat universe which is homogeneous and isotropic along with an appropriate equation of state (EOS). Let the scale factor be R with an energy density $\epsilon$, then the Friedman equation reads

$$\dot{R} - CR\sqrt{\epsilon} = 0 \qquad (1)$$

$$\dot{\epsilon} - 3(\dot{R}/R)(\epsilon + P) = 0$$

with C=$(8\pi/3)^{1/2}$/$M_p$ the Planck Mass $M_p$=1.2 x $10^9$ GeV. The corresponding equation of state, relating energy density $\epsilon$ and the pressure p using the bag model reads for QGP

$$\epsilon_{qg} = (37\pi^2/90)T^4 - B, \quad p_{qg} = (\epsilon_{qq} - 4B)/3 \tag{2a}$$

$$p = p_{qg} + p_{bg} : \epsilon = \epsilon_{qg} + \epsilon_{bg} : \epsilon_{bg} = p_{bg}/3 \tag{2b}$$

$$p_{bg} = 14.25\pi^2 T^4/90 \tag{2c}$$

The cosmic evolution will be an inflationary one if the expansion is accelerated, $\ddot{R} \geq 0$ which leads to using equation (1)

$$\ddot{R} = -C^2 R(\epsilon + p)/2 \geq 0$$

$$\text{and } 3p+\epsilon < 0 \tag{3}$$

with the solution

$$\epsilon = Bcth^2[2C\sqrt{B}(t - t_c) + arcth(\sqrt{\epsilon_c/B})] \tag{4a}$$

$$R = sh^{1/2}[2C\sqrt{B}(t - t_c) + arcth(\sqrt{\epsilon_c/B})sh^{-1/2}(arcth\sqrt{\epsilon_c/B})] \tag{4b}$$

at t≫ $t_{exp} = (2C\sqrt{B})^{-1}$ clearly the space expansion proceed exponentially, R ∝ $\exp(C\sqrt{B}t)$.

Equations 2 & 3 are satisfied for T<$T_i$ with $T_i \cong 0.5\ B^{1/B}$;

For temperature below $T_0$= 0.65$B^{1/4}$, the pressure becomes negative leading to acceleration of the universe. This is exactly what is achieved by the "mini inflation"

To recapitulate, the universe is assumed to begin with a large baryon chemical potential acquired through an Affleck- Dine [19] type of mechanism. It then undergoes a period of inflation, Fig 2 crossing the QCD first order phase transition line, while remaining in a deconfined and in a chirally symmetric phase. The universe is then trapped in a false metastable QCD vacuum state.

The delayed phase transition then releases the latent heat and produces concomitantly a large entropy density which effectively reduces the baryon asymmetry to currently observable values. It then enters a reheating phase all the way upto the usual reheating temperature with no significant change in the baryon potential and then the universe follows the standard path to lower temperatures.

Experimental observations of Alcock et al. [4], Aubourg et al. [5] and more recently by the Bose Institute group [13,14] are the clinching proof.

Finally, as suggested by Witten [18] stable quark lumps (nuggets) is extremely optimistic. It would be very lucky to be true. The relics as pointed out in this paper are the acid test of that luck. The recent observations [13,14] along with the old MACHO observations [4,5] seem to survive the acid test.

## 4. Survivability of Cosmological Strange Quark Nuggets (SQN)

In ref [11] a detailed discussion is presented about the survivability of SQN's tracing the history from Alcock and Farhi (see ref 25) onwards to Madson, Heisenberg and Riisager (see ref 11) and then to the detailed work of Bhattacharya et al. [8] using Chromo Electric Flux Tube (CEFT) came to the conclusion that QN's with baryon number $\geq 10^{39} - 10^{40}$ will indeed be cosmologically stable. It is thus very relevant to ask what fraction of the dark matter could be accounted for by the surviving QNs. To put it yet another way we wish to address in this paper is the proverbial cosmological dark matter, counting 90% or more of all the matter in the universe, can be made up entirely of QNs.

As per ref [9,10,11] the universe is closed by the baryonic dark matter trapped in QNs, we should have

$$N_B^H(t_p) = N_B^{QN} n_{QN} V^H(t_p) \tag{5}$$

where $N_B^H(t_p)$ is the total number of baryons required to close the universe ($\Omega_B = 1$) at $t_p$, $N_B^{QN}$ is the total number of baryons contained in a single QN, and $V^H(t_p) = (4\pi/3)(ct_p)^3$ is the horizon volume. With $v/c \approx 1/\sqrt{3}$

$$N_B^{QN} \leq 10^{-4.7} N_B^H(t_p) \tag{6}$$

As per standard Big Bang Nucleosynthesis (SBBN) $\eta \equiv n_B/n_\gamma$ ($10^{-9}$, $10^{-10}$) for convenience (as an estimate), the baryon number within the horizon at the QCD epoch $\approx 10^{49}$. This is rough estimate.

These usual baryons constitute only $\sim 10\%$ of the closure density ($\Omega_B \sim 0.1$ from SBBN), a total number of $10^{50}$ within the horizon at a temperature of $\sim 100$ MeV would close the universe baryonically, provided these baryons do not take part in SBBN, a criterion fulfilled by QNs. this would require $N_B^{QN} \leq 10^{45.3}$, clearly above the survivability limit of QNs.

5.  **Conclusion**

Driven by the familiar standard model and introducing a "mini inflation" at the cosmic quark hadron phase transition one can precipitate a first order phase transition from quarks to hadrons.

It has been argued in this paper that it does not introduce any non standard cosmological scenario and indeed this mini inflation come in quite naturally; raising the value of $\eta = n_B/n_\gamma$ substantially. It is argued that the MACHOs will survive the cosmological time scale and beyond a critical baryon number window of $\sim 10^{40}$ are the candidates of cold dark matter observed some years ago [4,5] and rather more recently [13,14].

The interestingly satisfying thought lingers that we can accommodate all this in the framework of standard model without invoking yet unobserved exotic physics.

**Acknowledgement:**

I like to thank my colleagues here, Partha Mazumdar, Debasis Majumdar, my young colleague Rudrajit Banerjee, Chiranjib Barman and of course Edward Witten for his insightful comments. The author would like to thank Indian National Science Academy (INSA) for the Emeritus position of INSA.

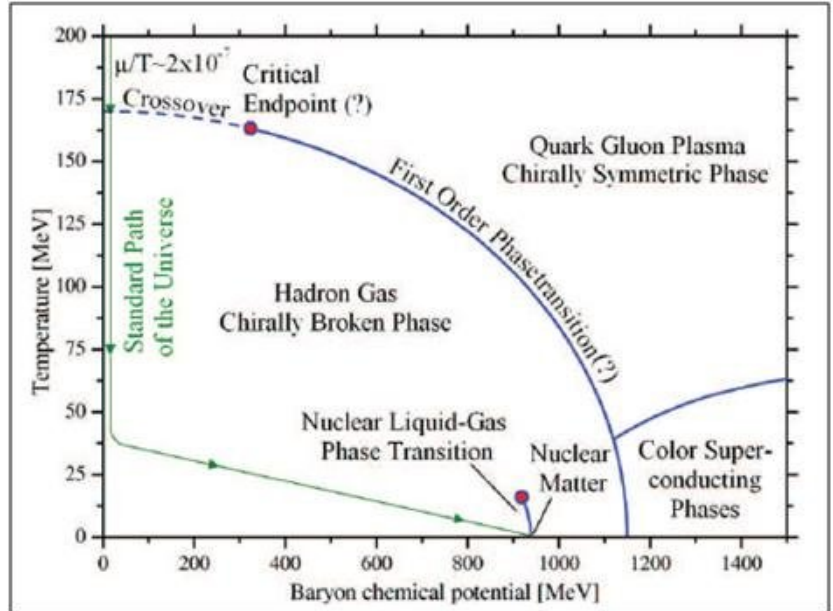

Fig 1: Sketch of a possible quantum chromodynamic phase diagram with the commonly accepted standard evolution path of the universe as calculated depicted by the green line.

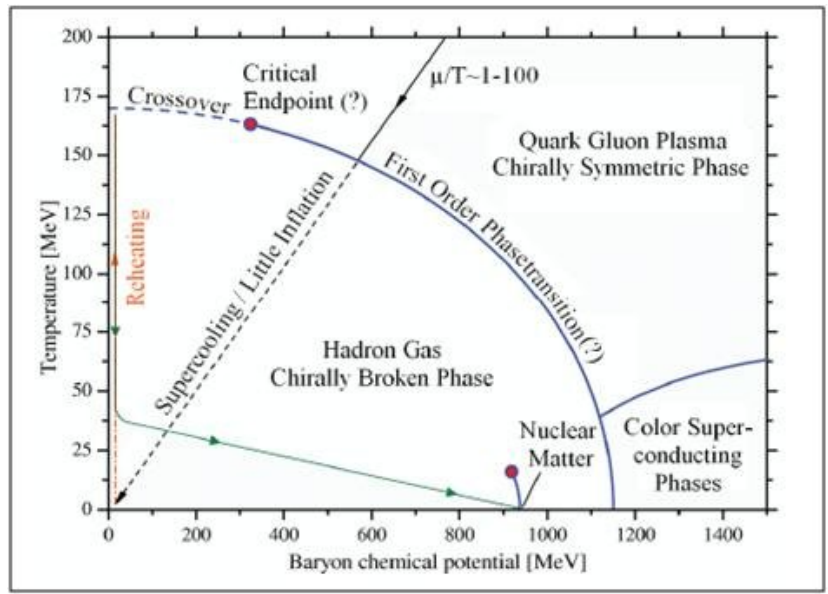

Fig 2: Sketch of a possible QCD phase diagram with the evolution path of the universe in the little inflation scenario.